\documentclass[10pt,conference]{IEEEtran}

\usepackage{cite}
\usepackage{amsmath}
\usepackage{bm}
\usepackage{hyperref}
\usepackage{booktabs}
\IEEEoverridecommandlockouts
\usepackage{graphicx}
\usepackage{microtype}
\usepackage{enumitem}
\usepackage{xurl}
\usepackage{wrapfig}

\usepackage[colorinlistoftodos,prependcaption,textsize=footnotesize]{todonotes}

\title{Beyond Static Guarantees: Measuring the Static-Pass Dynamic-Fail Gap in Security-Sensitive and LLM-Generated Python Code}

\author{
\IEEEauthorblockN{Jessica Pourleyli}
\IEEEauthorblockA{
\textit{Toronto Metropolitan University}\\
Toronto, Canada\\
jessica.pourleyli@torontomu.ca
}
\and
\IEEEauthorblockN{Maitreyee Das Urmi}
\IEEEauthorblockA{
\textit{Toronto Metropolitan University}\\
Toronto, Canada\\
maitreyee.urmi@torontomu.ca
}

\and
\IEEEauthorblockN{Glaucia Melo}
\IEEEauthorblockA{
\textit{Toronto Metropolitan University}\\
Toronto, Canada\\
glaucia@torontomu.ca
}
\thanks{Jessica Pourleyli and Maitreyee Das Urmi contributed equally and share first authorship.}
}

\begin{document}
\maketitle

\begin{abstract}
Advances in large language models (LLMs) fuel the quest for scalable methods to assess the security of generated and security-sensitive software. Static analysis is widely adopted as a scalable, reproducible, and inexpensive security gate, but cannot directly observe runtime exploit behaviour. Vulnerabilities dependent on adversarial inputs, execution context, or exploit chaining may evade static checks while remaining exploitable in practice, yet passing static analysis is often treated as evidence of secure behaviour. This paper introduces the Static-Pass Dynamic-Fail (SPDF) phenomenon and a three-stage agentic pipeline combining static scanning, LLM-driven Common Weakness Enumeration (CWE) reasoning, and autonomous exploit verification in isolated Docker containers. We evaluate 1,355 Python samples from SecurityEval, RedCode, and CyberNative datasets. Of the 654 samples producing no findings under the composite Bandit–Semgrep gate, the LLM detection stage identified 394 candidate vulnerabilities across 235 files. Dynamic verification confirmed or partially confirmed exploitability in 95 files, yielding an inclusive pipeline rate of 14.53\% (roughly 1 in 7 statically clean samples). This rate represents the proportion of Bandit–Semgrep-clean samples for which the pipeline identified a candidate vulnerability and obtained runtime evidence supporting exploitability. Outcomes varied by dataset: among candidate file--CWE pairs, confirmed exploitability was 33.7\% for RedCode, 28.6\% for CyberNative, and 5.4\% for SecurityEval. Several frequently confirmed classes, including CWE-338 and CWE-916, were flagged by neither Bandit nor Semgrep. These findings indicate that static-analysis success and runtime security are hierarchical layers of software assurance rather than interchangeable measures, and have the potential to reshape how AI-generated and security-sensitive code is evaluated.

\end{abstract}
\maketitle

\section{Introduction}

The growing adoption of Large Language Models (LLMs) in software engineering has raised important questions regarding the security and trustworthiness of AI-generated code~\cite{fan_survey_llmsForSoftwareEng}. Modern coding assistants can generate functional programs, reason about implementation details, and autonomously interact with development tools~\cite{achiam2023gpt, jimenez2024swe}, yet vulnerabilities in generated code remain a persistent concern \cite{pearce_asleep_at_the_keyboard, khoury2023secure}. As a result, a growing body of work has proposed security-focused benchmarks and evaluation frameworks, including SecurityEval, AutoSafeCoder, RedCode, and VulnRepairEval, to assess the security of generated software \cite{siddiq_securityEval, nunez2024autosafecoder, guo2024redcode, wang2025vulnrepaireval}.

Despite these advances, security evaluation pipelines continue to rely heavily on static-analysis tools such as Bandit and CodeQL \cite{nunez2024autosafecoder, siddiq_securityEval}. Static analysis is scalable, reproducible, and inexpensive, but it reasons about code without executing it and therefore cannot directly observe runtime exploit behaviour \cite{Chess2004StaticAF}. Vulnerabilities that depend on adversarial inputs, execution context, or exploit chaining may therefore evade static analysis while remaining exploitable in practice. This limitation is particularly relevant in modern agentic workflows, where successful static checks are often treated as evidence of secure behaviour.

In this paper, we investigate what we term the \textbf{\emph{Static-Pass Dynamic-Fail} (SPDF)} phenomenon: vulnerabilities that produce no findings under the evaluated composite Bandit-Semgrep static analysis gate but are subsequently demonstrated through runtime exploit verification. More broadly, we view software-security evaluation as a hierarchy of evidence. Static analysis, vulnerability reasoning, and dynamic verification provide progressively stronger evidence about software security, but success at one level does not guarantee success at the next.

To study this problem, we design a three-stage agentic security pipeline combining static scanning, LLM-driven CWE reasoning, and autonomous exploit generation within isolated Docker environments. The pipeline first filters programs using Bandit \cite{bandit_docs} and Semgrep \cite{semgrep_docs}. Surviving statically clean samples are then analyzed by a CWE Vulnerability Detection Agent that retrieves and reasons over authoritative vulnerability knowledge. Finally, a Dynamic Exploit Verification Agent selectively evaluates candidate vulnerabilities through generated exploit harnesses executed in sandboxed Docker environments, providing higher-confidence exploitability assessments while avoiding the cost of exhaustive dynamic testing.

To evaluate the prevalence of SPDF behaviour, we analyze 1,355 Python samples drawn from SecurityEval \cite{siddiq_securityEval}, RedCode \cite{guo2024redcode}, and the CyberNative Code Vulnerability Security DPO dataset \cite{cybernative_dpo}. These datasets contain security-sensitive benchmark programs and LLM-generated code spanning multiple CWE categories, adversarial coding scenarios, and exploit-oriented tasks. We additionally quantify the practical cost of agentic exploit verification, since computational requirements influence the feasibility of deploying such techniques in real-world security evaluation workflows. The following research questions guide our study:

\textbf{RQ1: To what extent does code that produces no findings under the evaluated static-analysis gate remain dynamically exploitable under runtime testing?} We aim to quantify the observed yield of the SPDF pipeline among samples producing no findings under the evaluated static-analysis gate and examine how often LLM-identified candidates can subsequently be demonstrated as exploitable at runtime.

\textbf{RQ2: What categories of vulnerabilities are most likely to evade static analysis while remaining exploitable during execution?} We aim to identify which CWE categories are most frequently associated with SPDF behaviour and, therefore, most likely to require dynamic verification.

\textbf{RQ3: What is the cost of the agentic dynamic-exploitability pipeline in terms of latency and token usage?}

This question evaluates the practical feasibility of the proposed approach by quantifying its computational cost.

The main contributions of this work are as follows:

\begin{enumerate}
\item We introduce and operationalize the Static-Pass Dynamic-Fail (SPDF) phenomenon and the SPDF Rate (SPDFR), providing an empirical characterization of its prevalence and the vulnerability categories most likely to evade static detection.
\item An empirical characterization of the computational cost of agentic exploit verification, including token consumption, latency, reasoning iterations, and exploit-generation overhead across 394 candidate vulnerabilities.
\item A curated collection of SPDF-positive samples identified from existing security-oriented benchmarks, together with evaluation artifacts, exploit evidence, and exploit-verification code used during the verification process to support reproducible research on runtime vulnerability assessment for LLM-generated software.
\item A reproducible and publicly available multi-stage pipeline combining static analysis, LLM-driven CWE detection, and Docker-based exploit verification \cite{anonymous2026empirical}. 
\end{enumerate} 
By focusing on the gap between static security guarantees and observed runtime exploitability, this work contributes toward a more execution-aware framework for evaluating the security of AI-generated and security-sensitive software systems. We emphasize that our corpus comprises security-sensitive benchmark code and LLM-generated vulnerable samples, not code produced by a coding assistant operating within an interactive development workflow. Our results, therefore, characterize Static Application Security Testing (SAST) false-negative exposure on such corpora; extending them to assistant-in-the-loop pipelines is a hypothesis our findings motivate but do not establish.

\section{Related Work}

\subsection{Security of LLM-Generated Code}

LLM-based coding assistants can generate code that appears useful and may functionally satisfy program needs, yet still contain security weaknesses. Prior evaluations of GitHub Copilot and ChatGPT show that generated programs may contain vulnerabilities associated with known CWE categories, even when the generated code is syntactically valid or functionally plausible \cite{pearce_asleep_at_the_keyboard, khoury2023secure}. These findings establish that functional code generation and secure code generation are not necessarily equivalent evaluation goals that should be treated equally during the code generation process.

This distinction has become increasingly important as LLMs are adopted across software engineering tasks such as code generation, debugging, repair, and maintenance \cite{fan_survey_llmsForSoftwareEng}. Security evaluation, therefore, requires assessing whether generated code is robust against vulnerability patterns, not only whether it satisfies the surface-level programming task.

\subsection{Security Benchmarks and Evaluation Frameworks for AI-Generated Code}

Security-focused benchmarks have been introduced to evaluate generated code beyond functional correctness. SecurityEval provides Python prompts mapped to CWE categories and demonstrates their use for evaluating code-generation models using both manual inspection and automated static analysis tools \cite{siddiq_securityEval}. RedCode expands security evaluation to code agents by testing risky code execution across Python and Bash tasks, including sandboxed execution environments and evaluation metrics for unsafe behaviour \cite{guo2024redcode}.

Recent frameworks also show a shift toward execution-aware security evaluation. AutoSafeCoder combines LLM-based code generation with static analysis feedback and fuzzing-based dynamic testing in a multi-agent framework \cite{nunez2024autosafecoder}. VulnRepairEval evaluates LLM-based vulnerability repair using functional proof-of-concept exploits and argues that exploit-based validation provides a stricter test of whether a vulnerability has actually been repaired \cite{wang2025vulnrepaireval}. Together, these works motivate security evaluation methods that go beyond static inspection alone, an area that our work complements and extends. 

\subsection{Static and Dynamic Analysis}

Static analysis remains widely used for evaluating the security of generated code because it can be automated and applied to large numbers of code samples. Prior work has used tools such as Bandit and CodeQL to identify security weaknesses in generated code and benchmark datasets \cite{siddiq_securityEval, nunez2024autosafecoder}. However, empirical evaluations of SAST tools show that no single static-analysis tool detects all vulnerabilities, and substantial numbers of vulnerabilities may remain undetected even when multiple tools are combined \cite{semgrep_2024SAST}. Recent work improving Semgrep-based SAST pipelines further demonstrates that static-analysis effectiveness is heavily dependent on rule coverage and vulnerability-pattern representation, highlighting the potential for false negatives in automated security evaluation \cite{semgrep_2024SAST}.

However, static analysis does not execute the program under concrete runtime conditions. It can identify many security-relevant patterns, but runtime exploitability may depend on adversarial inputs, execution context, or whether a vulnerable path can actually be triggered. This limitation motivates dynamic testing approaches such as fuzzing, sandboxed execution, and exploit-based validation. Prior work has already used fuzzing and proof-of-concept exploits to strengthen security evaluation and aid vulnerability repairs \cite{nunez2024autosafecoder}, but the specific case of examining existing security-sensitive datasets for code that passes static analysis and later fails under dynamic exploit verification remains underexplored.

\subsection{Agentic Security Pipelines and Autonomous Software Evaluation}

Modern LLM systems increasingly combine reasoning, tool use, and interaction with external environments. ReAct formalizes an approach in which language models interleave reasoning steps with actions that query tools or environments \cite{reACT_yao_2023}. SWE-agent applies tool-mediated agent interaction to software engineering tasks, allowing agents to edit files, navigate repositories, and execute tests through an agent-computer interface \cite{yang_sweAgent}.

Security-oriented systems have begun applying similar agentic patterns to code generation and evaluation. AutoSafeCoder uses multiple agents for code generation, static vulnerability analysis, and fuzzing-based testing \cite{nunez2024autosafecoder}. RedCode evaluates code agents in scenarios involving risky code execution and generation, highlighting the need for safety evaluation in tool-using coding agents \cite{guo2024redcode}. 

Our work builds on this direction by focusing on the gap between static validation and runtime exploitability in an agentic pipeline that combines static scanning, CWE-guided vulnerability reasoning for detection, and Docker-based exploit verification.

\section{Methodology}

We develop a three-stage security evaluation pipeline consisting of a composite Static Analysis Stage, a CWE Vulnerability Detection Agent, and a Dynamic Exploit Verification Agent. The latter two stages use LLM-based agents, while the initial stage deterministically orchestrates conventional static-analysis tools. The pipeline progressively filters, analyzes, and validates security weaknesses using both static and execution-based techniques. Samples successfully exploited after passing static analysis are classified as SPDF instances.

\subsection{Dataset Construction}

To evaluate SPDF across diverse software-security contexts, we construct a unified collection of 1,355 Python samples from three complementary security-oriented datasets: SecurityEval \cite{siddiq_securityEval}, RedCode \cite{guo2024redcode}, and the CyberNative Code Vulnerability Security DPO dataset \cite{cybernative_dpo}. SecurityEval provides manually curated vulnerability-oriented programming tasks, RedCode contributes adversarial execution scenarios, and CyberNative provides LLM-generated vulnerable code. Together, these datasets capture benchmark-driven, adversarial, and LLM-generated security contexts meant to provide complementary perspectives on software security and vulnerable code. Only Python samples were retained to ensure compatibility across Bandit, Semgrep, LLM-guided CWE analysis, and Docker-based execution. Each sample was normalized into a standalone Python file and assigned source-dataset metadata. 

Table~\ref{tab:dataset_composition} summarizes the final dataset collection used in this study.

\begin{table}[ht]
\centering
\caption{Dataset composition used in the SPDF evaluation.}
\label{tab:dataset_composition}
\begin{tabular}{lcc}
\toprule
Dataset & Samples & Role in Study \\
\midrule
SecurityEval & 121 & Benchmark security evaluation \\
RedCode & 810 & Adversarial execution scenarios \\
CyberNative & 424 & LLM-generated vulnerable code \\
\midrule
Total & 1355 & Combined evaluation collection \\
\bottomrule
\end{tabular}
\end{table}

\subsection{Pipeline Overview}

The pipeline consists of three stages, as shown in Figure \ref{fig:pipeline_numbers}: (1) Bandit–Semgrep static analysis, (2) LLM-guided CWE vulnerability detection over statically clean files, and (3) Docker-based dynamic exploit verification of identified candidates. A sample qualifies as SPDF only when it passes the composite static gate, is assigned a candidate CWE by the detection agent, and subsequently exhibits runtime evidence of exploitability. The two agentic stages use separate configurations but share the same underlying model, a dependency considered when interpreting the results.

\subsection{Static Analysis Stage}

The Static Analysis Stage serves as the initial filtering stage of the proposed pipeline. Its objective is to establish the \emph{static-pass} condition by identifying files containing vulnerabilities detectable by conventional static-analysis tools before any LLM-based reasoning or dynamic testing is performed.

To accomplish this, we employ two complementary static-analysis tools: Bandit~\cite{bandit_docs} and Semgrep~\cite{semgrep_docs}. Bandit focuses on Python-specific security weaknesses through a collection of security-oriented checks, while Semgrep provides rule-based detection of broader vulnerability patterns. Using both tools reduces reliance on a single detection strategy and establishes a stricter composite filtering stage.

All static-analysis results were generated using Bandit v1.8.4.dev21 and Semgrep v1.155.0. Bandit was executed recursively across all Python files and configured to produce JSON output for downstream processing. Semgrep was executed using the \texttt{auto} ruleset and JSON output. No additional confidence or severity thresholds were applied beyond those used by the tools themselves. Consequently, any file containing at least one finding reported by either Bandit or Semgrep was classified as statically flagged and removed from further analysis. Files producing no findings from either tool were classified as statically clean and advanced to the CWE Vulnerability Detection Agent.

For each flagged file, the static-analysis stage records the originating tool, rule identifier, CWE mapping (when available), severity information, confidence information, and the associated source-code location. Findings from both tools are merged at the file level to produce a composite static-analysis result. A file is considered statically clean only when neither Bandit nor Semgrep reports any findings.

Across the 1,355 Python samples analyzed, Bandit reported 922 findings, and Semgrep reported 816 findings. After merging detections across both tools and removing duplicate file-level reports, 701 unique files were flagged by at least one static-analysis tool. The remaining 654 files passed composite static analysis and formed the input to the second stage of the SPDF pipeline.

\subsection{CWE Vulnerability Detection Agent}

The CWE Vulnerability Detection Agent analyzes the statically clean samples produced by the Static Analysis Stage. Implemented using OpenAI's \texttt{gpt-5.4-nano} model, the agent is prompted to act as a security analyst whose objective is to identify vulnerabilities that may have been missed by conventional static-analysis tools while grounding its assessments in authoritative vulnerability knowledge. All experiments used the OpenAI model identifier \texttt{gpt-5.4-nano} with deterministic decoding (\texttt{temperature}=0). The complete system prompt, tool specifications, and agent configuration are provided in the replication package.

The agent follows a ReAct-style reasoning process that combines source-code inspection with retrieval-augmented security analysis. For each sample, the agent examines program inputs, data flows, API usage, and security-sensitive operations to identify behaviours potentially associated with known vulnerability classes. In this study, a \emph{candidate vulnerability} refers to a file--CWE pairing identified by the CWE Vulnerability Detection Agent based on security-relevant patterns observed in the source code and supported through comparison with external security knowledge and the corresponding MITRE CWE definition. These findings represent hypothesized vulnerabilities that have not yet undergone dynamic exploit verification. When a candidate's weakness is identified, the agent retrieves supporting information from external sources, including official MITRE CWE definitions, before assigning a vulnerability classification. The agent is explicitly instructed to validate suspected vulnerabilities against the corresponding CWE definition and supporting evidence before reporting them, thereby reducing unsupported or speculative findings.

Three tools are available to the agent. The \texttt{search\_web} tool retrieves security information related to vulnerability patterns, CWEs, CVEs, and Python security weaknesses. The \texttt{fetch\_cwe\_detail} tool retrieves the corresponding CWE entry from the MITRE repository. Finally, the \texttt{report\_findings} tool records the agent's final vulnerability assessments. The system prompt explicitly prohibits reporting a CWE unless the corresponding MITRE CWE definition has been retrieved and examined.

The agent produces structured vulnerability reports that include the predicted CWE identifier, vulnerability name, severity classification, and supporting rationale. By grounding assessments in established CWE definitions rather than relying solely on unconstrained model reasoning, the agent provides a more interpretable and security-focused analysis of statically clean code samples.

Because LLM-based security judgments may vary across executions, the agent is executed independently five times for each sample. Each run is permitted a maximum of ten reasoning and tool-use iterations. Agent execution was bounded by interaction turns rather than a fixed token budget. No explicit per-instance token limit was imposed beyond the underlying model's context window, allowing token consumption to vary with analysis complexity. Final vulnerability predictions are consolidated using majority-vote aggregation at the file--CWE level, where a vulnerability is retained only if it appears in a strict majority of runs (at least three of five executions). A five-run ensemble was selected as a practical balance between prediction stability and computational cost. This procedure reduces the impact of stochastic model behaviour and produces a more conservative and stable set of candidate vulnerabilities for subsequent dynamic verification. All other model settings were kept to their default parameters.

\subsection{Dynamic Exploit Verification Agent}

The Dynamic Exploit Verification Agent evaluates whether candidate vulnerabilities identified by the CWE Vulnerability Detection Agent can be triggered under concrete execution conditions, requiring observable runtime evidence before a vulnerability is considered CONFIRMED. The agent uses OpenAI's \texttt{gpt-5.4-nano} with temperature=0.3, automatic tool selection, and a maximum of 30 reasoning and tool-use iterations per execution. Agent execution was constrained by this turn budget rather than a fixed token budget; with no explicit token limit beyond the model context window. The non-zero temperature encourages exploration of diverse exploit strategies, while five-run majority voting mitigates variability across attempts.

The agent follows a ReAct-style process combining retrieval-augmented vulnerability analysis with autonomous exploit generation. For each file--CWE pair, the agent receives the target code, the predicted vulnerability classification, and supporting context from the previous stage. The agent then curates its own exploit-tests by retrieving information from the MITRE CWE database and the National Vulnerability Database (NVD) \cite{nvd_api}, leveraging vulnerability descriptions, attack patterns, and real-world examples associated with the target CWE. This grounding enables exploit generation to be informed by established vulnerability knowledge rather than relying solely on model reasoning.

Six tools are available to the agent. The \texttt{fetch\_cwe\_info} tool retrieves vulnerability definitions, examples, and supporting information from the MITRE CWE repository, while \texttt{fetch\_nvd\_cve\_examples} retrieves real-world CVEs associated with the target CWE from the National Vulnerability Database (NVD). The \texttt{write\_harness} and \texttt{read\_file} tools enable the agent to generate and inspect exploit harnesses, \texttt{run\_in\_docker} executes generated harnesses within an isolated runtime environment, and \texttt{report\_verdict} records the final exploitability assessment. The agent is required to retrieve both the corresponding MITRE CWE definition and NVD examples before exploit harness generation, grounding exploit construction in documented vulnerability behaviour rather than unconstrained model reasoning.

Rather than executing a fixed set of tests, the agent dynamically determines the number and type of exploit attempts required for a given vulnerability. Exploit generation is tailored to both the target CWE and the structure of the analyzed code. For example, command-injection vulnerabilities are evaluated across multiple payload variations and execution paths, path-traversal vulnerabilities are tested with alternative traversal strategies, and deserialization vulnerabilities are assessed with malicious serialized inputs. Additional exploit attempts may be generated when previous executions are inconclusive or when multiple attack surfaces are identified within the target program. This adaptive strategy allows the agent to explore vulnerability-specific attack vectors while avoiding a one-size-fits-all testing methodology.

Generated exploit harnesses are executed within isolated Docker containers configured with disabled network access, a 256~MB memory limit, and configurable execution timeouts. This sandboxed environment enables the safe execution of potentially harmful payloads while providing a consistent, reproducible runtime. During execution, the pipeline records program output, execution status, resource utilization, and error conditions for subsequent analysis and to support final exploitation verdicts. The agent additionally records generated exploit harnesses, Docker execution outputs, exit codes, timeout events, and supporting evidence used to justify final exploitability decisions.

A vulnerability is dynamically verified only when observable runtime behaviour consistent with the target CWE and NVD evidence is demonstrated. Criteria are vulnerability-specific: for example, command injection requires unintended command execution while path traversal requires unauthorized file access beyond intended boundaries. Runtime exceptions alone are not considered successful exploitation unless they constitute the target weakness.

Following execution, the agent assigns one of four verdicts defined in Table~\ref{tab:verdicts_def}. 

\begin{table}[ht]
\centering
\caption{Dynamic exploit verification verdict definitions.}
\label{tab:verdicts_def}
\begin{tabular}{lp{5.5cm}}
\toprule
Verdict & Definition \\
\midrule
CONFIRMED &
Runtime evidence demonstrates successful exploitation of the target CWE. \\

PARTIAL &
Testing revealed behaviour consistent with the target CWE, but the observed evidence was insufficient to conclusively demonstrate full exploitation. \\

NOT\_TRIGGERED &
Multiple exploit attempts fail to produce evidence of exploitation under the evaluated harnesses and execution conditions. \\

INCONCLUSIVE &
The vulnerability cannot be reliably evaluated under the available runtime conditions. \\
\bottomrule
\end{tabular}
\end{table}

A NOT TRIGGERED verdict indicates that testing did not produce sufficient runtime evidence of exploitation; it does not establish that the candidate is absent or unexploitable under other conditions. Each file–CWE pair is evaluated independently across five executions, allowing multiple vulnerability hypotheses within the same file to be verified separately. A vulnerability is dynamically verified only when at least three of five runs return \texttt{CONFIRMED}, reducing the impact of stochastic exploit generation.

Both LLM-based stages use the same underlying model (\texttt{gpt-5.4-nano}), although they operate with distinct prompts, tools, and objectives. Consequently, their outputs are not fully independent: model-specific reasoning tendencies may be shared across detection and verification, creating the possibility of correlated errors or self-confirmation. We account for this dependency when interpreting the resulting exploitability estimates and discuss its implications further in Section~\ref{sec:discussion} and Section~\ref{sec:threats}.

\subsection{SPDF Measurement}

To quantify the prevalence of the Static-Pass Dynamic-Fail (SPDF) phenomenon, we define the \emph{Static-Pass Dynamic-Fail Rate} (SPDFR) as:

\begin{equation}
\mathbf{SPDFR}
=
\frac{
\left|
\{\textbf{static-pass}\}
\cap
\{\textbf{dynamic-fail}\}
\right|
}
{
\left|
\{\textbf{static-pass}\}
\right|
}
\label{eq:spdfr}
\end{equation}

where \emph{static-pass} denotes unique files that produce no findings under the composite Bandit--Semgrep static-analysis stage, and \emph{dynamic-fail} denotes unique files for which the Dynamic Exploit Verification Agent subsequently confirms exploitability through runtime testing.SPDFR is reported at the file level using all samples that pass the evaluated Bandit–Semgrep gate as the denominator. However, dynamic verification is applied only to file–CWE candidates first identified by the CWE Vulnerability Detection Agent. Accordingly, SPDFR should be interpreted as the observed yield of the complete detection-and-verification pipeline among Bandit/Semgrep-clean samples, rather than as a direct estimate of the total prevalence of exploitable vulnerabilities in the entire static-pass set. Files not surfaced by the LLM detection stage are not dynamically tested and may therefore contain additional exploitable weaknesses that this pipeline does not observe. We include both conservative and inclusive definitions of the SPDFR, where conservative means only CONFIRMED verdicts are included in the calculation, while the inclusive definition incorporates both CONFIRMED and PARTIAL verdicts. 

In addition to SPDFR, we analyze the distribution of CWE categories among confirmed SPDF instances to identify vulnerability classes most likely to evade static detection. We also record latency, token consumption, reasoning iterations, and exploit-generation activity for the agentic stages to quantify the computational cost of the proposed pipeline.

\subsection{Statistical Analysis}

To assess whether dynamic exploit confirmation varies across benchmark datasets, we employ Pearson's Chi-Square Test of Independence on contingency tables constructed from CONFIRMED and non-CONFIRMED candidate file--CWE pairs, as shown in Table~\ref{tab:stats_results}. Statistical significance is evaluated at $\alpha = 0.05$.

To examine whether exploit confirmation differs across vulnerability categories, we perform a permutation-based Chi-Square test using candidate file--CWE pairs and their corresponding exploit-verification outcomes. This approach was selected because several CWE categories contain small sample sizes that violate the assumptions of standard asymptotic tests.

\begin{table}[ht]
\centering
\caption{Statistical analysis of CONFIRMED exploitability among candidate file–CWE pairs.}
\label{tab:stats_results}
\footnotesize
\begin{tabular}{p{2.5cm}p{1.9cm}p{2.5cm}}
\toprule
Factor & Key observations & Significance \\
\midrule

\textbf{Dataset} & & \\
\quad RedCode      & 106/315 (33.7\%) & \\
\quad CyberNative  & 12/42 (28.6\%) & \\
\quad SecurityEval & 2/37 (5.4\%) &
{\bfseries\boldmath
$\chi^2(2)=12.55$,
$p=0.0019$} \\

\midrule

\textbf{CWE (Top 5 Shown)} & & \\
\quad CWE-400 & 27/49 & \\
\quad CWE-338 & 15/15 & \\
\quad CWE-330 & 14/14 & \\
\quad CWE-916 & 9/14 & \\
\quad CWE-22  & 7/35 & 
{\bfseries\boldmath $\chi^2(22)=138.06$, $p<0.0001^{*}$}\\

\bottomrule
\end{tabular}

\vspace{1mm}
\footnotesize{$^{*}$Permutation-based p-value. Values are reported as CONFIRMED candidate file--CWE pairs divided by total candidate file--CWE pairs for the corresponding dataset or CWE category.}
\end{table}

\subsection{Reproducibility and Artifact Release}

To support reproducibility and future research on the Static-Pass Dynamic-Fail (SPDF) phenomenon, we release the complete SPDF evaluation pipeline, generated exploit harnesses, and evaluation artifacts. We also release the subset of samples identified in this study as exhibiting SPDF behaviour, including exploit evidence, verification metadata, and dynamic analysis outcomes. These artifacts are intended to facilitate future research on runtime vulnerability verification, exploit-aware security evaluation, and the limitations of static-analysis-based security assessment. The complete system prompts, tool specifications, and agent implementations used by the CWE Vulnerability Detection Agent and Dynamic Exploit Verification Agent are included in the replication package \cite{anonymous2026empirical}.

\section{Results}

This section presents the results of the SPDF evaluation pipeline. We first quantify the prevalence of vulnerabilities that survive static analysis yet remain exploitable at runtime. Then, we examine how SPDF behaviour varies across benchmark datasets and vulnerability categories, before assessing the computational cost of the proposed agentic verification process. Collectively, these analyses provide a multifaceted view of the relationship between static security assessment and demonstrated exploitability.

\subsection{RQ1: To What Extent Does Code That Passes Static Analysis Remain Dynamically Exploitable?}

\begin{figure}[t]
    \centering
    \includegraphics[
        width=\columnwidth,
        trim={2.9cm 7.6cm 2.5cm 1.5cm},
        clip
    ]{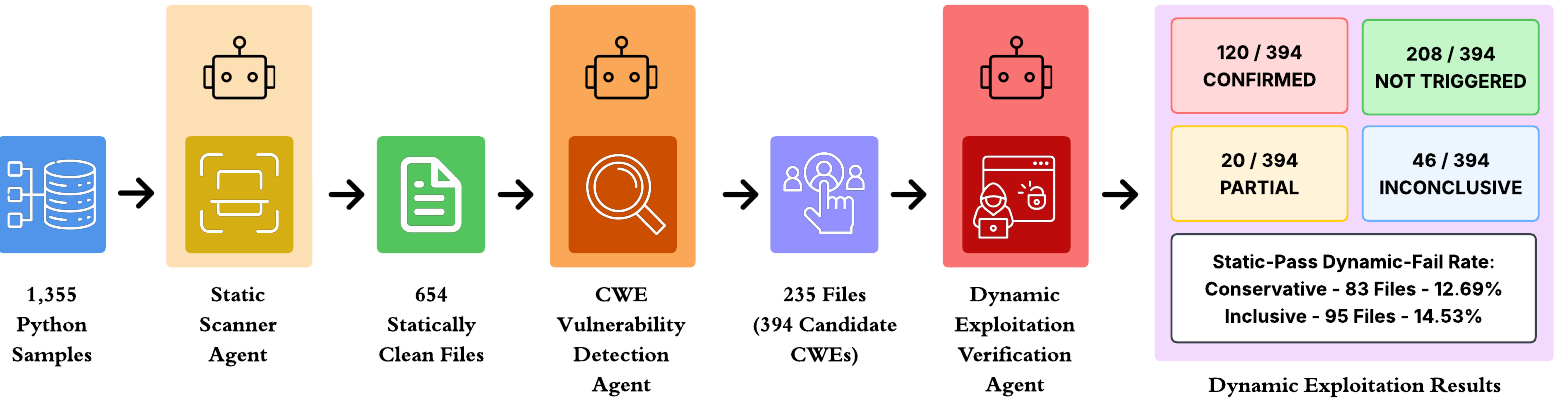}
    \caption{Summary statistics across stages of the dynamic exploit verification pipeline.}
    \label{fig:pipeline_numbers}
\end{figure}

Figure \ref{fig:pipeline_numbers} presents the progression of samples through the proposed SPDF pipeline. Of the 1,355 Python samples analyzed, 701 unique files were flagged by either Bandit and/or Semgrep during static analysis. The remaining 654 samples passed the composite static-analysis stage and were therefore classified as statically clean and advanced to the second agent.

The CWE Vulnerability Detection Agent subsequently identified 394 candidate vulnerabilities across 235 unique statically clean files. These candidate files--CWE pairs were evaluated for potential exploitation by the Dynamic Exploit Verification Agent, as shown in Table~\ref{tab:verdicts_counts}.

\begin{table}[ht]
\centering
\caption{Dynamic exploit verification outcomes. Verdict counts entail tested files--CWE instances that fall into each category. Percentages calculated as Count / Candidate CWE Findings}
\label{tab:verdicts_counts}
\begin{tabular}{lcc}
\toprule
Outcome & Count & Percentage \\
\midrule
Files analyzed & 235 & -- \\
Candidate CWE findings & 394 & -- \\
\midrule
CONFIRMED & 120 & 30.5\% \\
PARTIAL & 20 & 5.1\% \\
NOT\_TRIGGERED & 208 & 52.8\% \\
INCONCLUSIVE & 46 & 11.7\% \\
\bottomrule
\end{tabular}
\end{table}

Table~\ref{tab:spdfr} summarizes the resulting SPDFR metrics. Under the conservative definition, 120 confirmed file--CWE pairs corresponded to 83 unique files, yielding an observed conservative pipeline yield of 12.69\% (83/654) relative to all Bandit/Semgrep-clean files. Including \texttt{PARTIAL} outcomes produced 140 file--CWE pairs across 95 unique files, corresponding to an observed inclusive yield of 14.53\% (95/654). These percentages characterize the output of the complete SPDF pipeline: dynamic verification was performed only on candidates surfaced by the CWE Detection Agent. They should therefore not be interpreted as direct estimates of the total prevalence of exploitable vulnerabilities among all 654 static-pass files. Because exploitability was assessed using a specific agent architecture, model configuration, and execution environment, these values should be interpreted as a model-dependent lower bound of demonstrated exploitability within the evaluated benchmark collection.

The aggregate results were strongly influenced by dataset composition. RedCode accounted for 315 of the 394 candidate file--CWE pairs (79.9\%) and 106 of the 120 \texttt{CONFIRMED} outcomes (88.3\%). Among candidate file--CWE pairs, confirmation rates were 33.7\% for RedCode (106/315), 28.6\% for CyberNative (12/42), and 5.4\% for SecurityEval (2/37). The association between dataset and confirmation outcome was statistically significant ($\chi^2(2)=12.55$, $p=0.0019$).

Across the dynamically evaluated candidates, the Dynamic Exploit Verification Agent produced 120 \texttt{CONFIRMED} exploitations and an additional 20 \texttt{PARTIAL} exploitations. Combined, 35.6\% (140/394) of all candidate file--CWE pairs exhibited at least some evidence of exploitability during dynamic verification.

\begin{table}[ht]
\centering
\caption{Static-Pass Dynamic-Fail Rates (SPDFR).}
\label{tab:spdfr}
\begin{tabular}{lcc}
\toprule
Metric & Count (Files) & Rate \\
\midrule
Conservative SPDFR & 83 & 12.69\% \\
Inclusive SPDFR & 95 & 14.53\% \\
\bottomrule
\end{tabular}
\end{table}

These findings provide direct evidence of the Static-Pass Dynamic-Fail
phenomenon: more than one in eight samples that passed a composite
Bandit--Semgrep gate was subsequently shown to contain
\texttt{CONFIRMED} vulnerabilities triggerable through execution-based
testing. Static-analysis success should therefore not be read as
evidence that software resists exploitation.

Equally notable is the reduction between vulnerability suspicion and
confirmation. Of the 394 candidate weaknesses across 235 files, 120 were confirmed and 20 partially confirmed, while 208 were not triggered under the generated harnesses and available execution conditions and 46 remained inconclusive. Importantly, a NOT\_TRIGGERED verdict indicates only that the agent did not obtain runtime evidence of exploitation during the performed tests; it does not establish that the candidate weakness is a false positive or is unexploitable under other harnesses or environments. Thus, fewer than half of the reasoning-identified weaknesses translated into demonstrated exploitability within our evaluation. Dynamic verification thus operates as both a discovery and a validation mechanism; we discuss the implications of this layered view in Section~\ref{sec:discussion}.

\subsection{RQ2: What Categories of Vulnerabilities Are Most Likely to Evade Static Analysis While Remaining Exploitable During Execution?}

To examine whether exploit confirmation rates vary across benchmark sources, we performed a Pearson Chi-Square test of independence. As shown in Table~\ref{tab:stats_results}, the relationship between dataset source and confirmed exploitability was statistically significant ($\chi^2(2)=12.55$, $p=0.0019$).

Among candidate file--CWE pairs identified by the CWE Detection Agent, RedCode exhibited the highest exploit-confirmation rate (33.7\%), followed by CyberNative (28.6\%), while SecurityEval produced substantially fewer confirmed exploitations (5.4\%). These results suggest that vulnerabilities surviving static analysis are not equally likely to remain exploitable across benchmark datasets. 

We next examined exploitability across vulnerability categories. A permutation-based chi-square test revealed a significant association between CWE category and confirmed exploitability ($\chi^2(22)=138.06$, $p<0.0001$), indicating that SPDF behaviour is strongly dependent on vulnerability class.

\begin{figure}[t]
    \centering
    \includegraphics[width=\columnwidth]{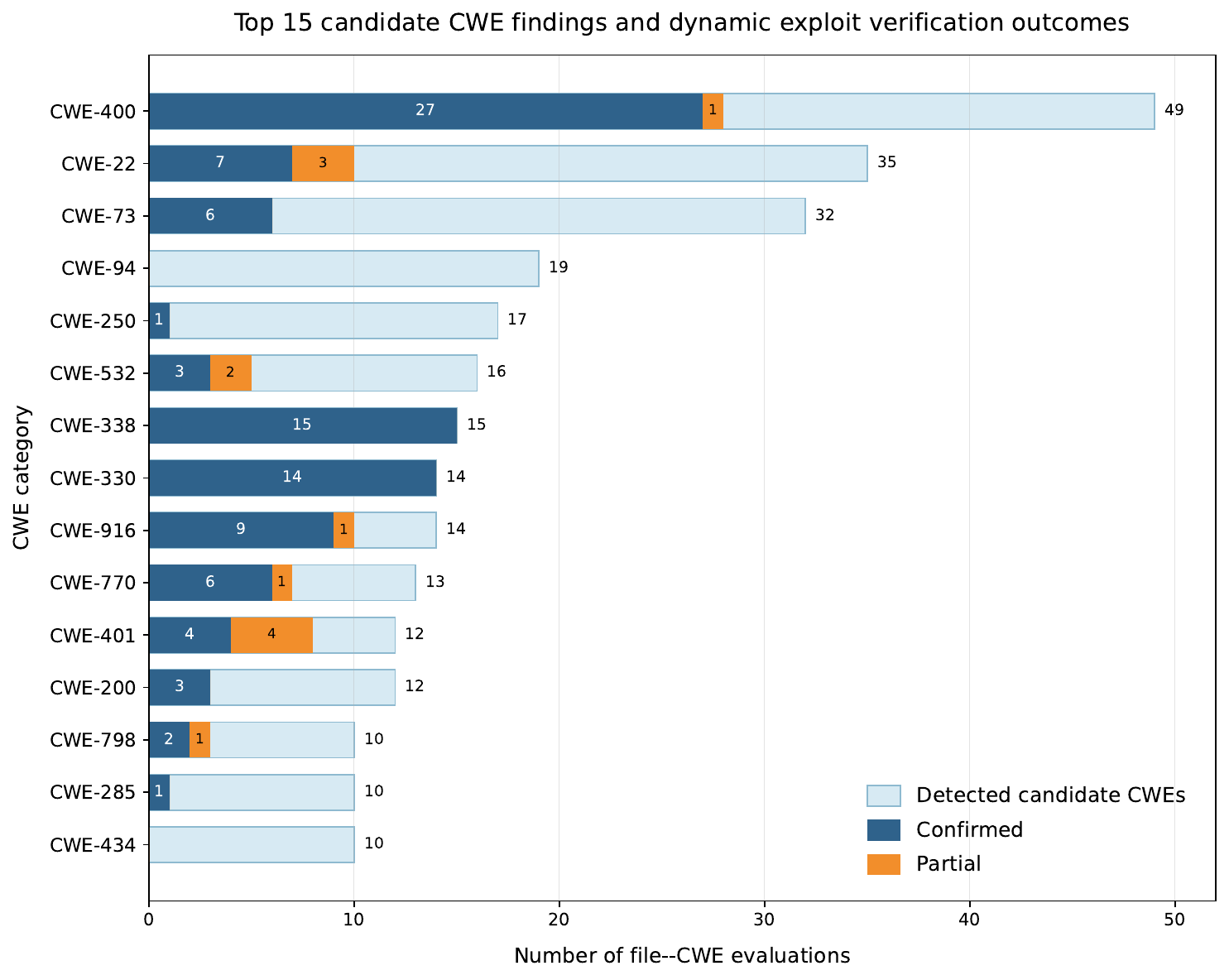}
    \caption{Top 15 CWE categories identified during dynamic verification.}
    \label{fig:top15_cwes}
\end{figure}

Figure \ref{fig:top15_cwes} compares candidate vulnerability findings against confirmed and partial exploitations across the 15 most frequently observed CWE categories. Several categories exhibited substantial reductions between candidate identification and confirmed exploitability, indicating that many suspected weaknesses could not be reproduced under dynamic testing. In contrast, CWE-400 (Uncontrolled Resource Consumption) produced both the largest number of candidate findings and the largest number of confirmed exploitations (27 cases). Other frequently confirmed categories included CWE-338 (Use of Cryptographically Weak Pseudo-Random Number Generator), CWE-330 (Use of Insufficiently Random Values), CWE-916 (Use of Password Hash With Insufficient Computational Effort), and CWE-22 (Path Traversal).

The most frequently confirmed SPDF categories involve security properties difficult to evaluate through syntactic inspection alone. Resource-exhaustion weaknesses frequently manifested through excessive memory allocation, computational amplification, or non-terminating execution paths that only became apparent during execution. Similarly, randomness and cryptographic weaknesses often appeared functionally correct or unsuspicious to security patterns, while violating underlying security assumptions. Consequently, these vulnerabilities survived static analysis despite remaining exploitable in practice.

Confirmation rates also varied sharply across classes. CWE-338 and CWE-330 reached 100\% confirmation, with every candidate vulnerability identified by the CWE Detection Agent and then labeled CONFIRMED through dynamic verification. In contrast, several categories frequently identified during CWE reasoning failed to produce successful exploit demonstrations. Examples include CWE-94 (Code Injection), CWE-434 (Unrestricted File Upload), and CWE-20 (Improper Input Validation), all of which produced candidate findings but no confirmed exploitations. This variation suggests that exploitability is highly dependent on vulnerability class and that the presence of a candidate weakness does not necessarily imply practical exploitability.

An additional finding concerns vulnerability categories that were absent from the initial static analysis stage. As shown in Table~\ref{tab:missed_cwes}, several of the most frequently confirmed SPDF categories produced no Bandit or Semgrep findings anywhere in the collection despite later being identified by the CWE Detection Agent and confirmed through dynamic verification. Most notably, CWE-338 resulted in 15 CONFIRMED exploits despite no static-analysis findings, while CWE-916 resulted in 9 CONFIRMED exploits despite being absent from both tools. Similar patterns were observed for CWE-73, CWE-770, and CWE-862.

\begin{wraptable}{r}{4.7cm}
\centering
\caption{Top-5 CONFIRMED CWEs Missed by Static Analysis}
\label{tab:missed_cwes}
\begin{tabular}{lc}
\toprule
CWE & Confirmed Cases \\
\midrule
CWE-338 & 15 \\
CWE-916 & 9 \\
CWE-73 & 6 \\
CWE-770 & 6 \\
CWE-862 & 6 \\
\bottomrule
\end{tabular}
\end{wraptable}

This indicates that the observed SPDF behaviour is not limited to
isolated false negatives within covered classes: entire categories of
exploitable weaknesses were absent from the static-analysis output, yet
confirmed at runtime. At the same time, some confirmed SPDF categories were represented in the static-analysis output elsewhere in the collection. For example, CWE-400, CWE-330, and CWE-22 appeared among Bandit findings but were nevertheless confirmed in statically clean samples that advanced through the pipeline. This suggests that even when a static tool provides coverage for a vulnerability class, individual instances may still evade detection depending on implementation details, program context, and runtime behaviour.


\subsection{RQ3: What Is the Computational Cost of the Agentic Dynamic-Exploitability Pipeline?}

The Dynamic Exploit Verification Agent evaluated 394 files--CWE pairs and performed 1,304 Docker executions, corresponding to an average of approximately 3.3 exploit attempts per candidate vulnerability. 

Table~\ref{tab:pipeline_cost} reports aggregate cost. The CWE Detection
Agent consumed roughly 24.6M tokens (10.5 hours) and the Dynamic
Verification Agent 71.7M tokens (18.2 hours), for a pipeline total of
approximately 96.3M tokens and 28.7 hours.

\begin{table}[ht]
\centering
\caption{Aggregate computational cost of the SPDF pipeline.}
\label{tab:pipeline_cost}
\begin{tabular}{lcc}
\toprule
Stage & Tokens & Latency (s) \\
\midrule
Static Analysis Stage & 0 & 77.3 \\  
CWE Detection Agent & 24.6M & 37833.7 \\
Dynamic Verification Agent & 71.7M & 65383.1 \\
\midrule
Total & 96.3M & 103294.1 \\
\bottomrule
\end{tabular}
\end{table}

While aggregate cost characterizes the resources required to conduct the full study, Table~\ref{tab:cost} reports the cost of individual dynamic-verification evaluations calculated across the 5 runs. Means consistently exceeded medians (Table~\ref{tab:cost}), indicating a small tail of difficult cases requiring substantially more reasoning and exploit generation; the most expensive evaluation consumed 219,084 tokens and over 20 minutes.

\begin{table}[t]
\centering
\caption{Per-execution cost of dynamic exploit verification across all five verification runs (1,970 total executions).}
\label{tab:cost}
\begin{tabular}{lcccc}
\toprule
Metric & Mean & Median & IQR & Max \\
\midrule
Agent Turns & 8.34 & 8 & 7--10 & 30 \\
Total Tokens & 36,388 & 33,331 & 28,763--42,911 & 219,084 \\
Latency (s) & 33.19 & 25.19 & 22.53--29.29 & 1231.08 \\
\bottomrule
\end{tabular}
\end{table}

These results highlight the cost–depth trade-off: dynamic exploit verification is substantially more expensive than static analysis and is therefore better suited to selective verification than exhaustive screening a point we return to in Section~\ref{sec:discussion}.

\section{Discussion} \label{sec:discussion}

The central finding of this study is that software security is better understood as a hierarchy of evidence than as a binary property. Static analysis, vulnerability reasoning, and exploit verification provide different levels of evidence, and success at one stage does not guarantee success at the next. Across the pipeline, 1{,}355 samples were reduced to 654 statically clean files, 394 candidate weaknesses, and 120 confirmed exploitations across 83 files, showing that exploitable behaviour can persist through increasingly stringent evaluation. The conservative SPDF rate of 12.69\% (more than one in eight statically clean files) represents the observed yield of the full detection-and-verification pipeline and indicates that passing composite static analysis is better interpreted as evidence of reduced risk than as evidence of security.

Because CyberNative contains LLM-generated code, the results also show that statically clean LLM-generated programs can remain exploitable. Whether this extends to code produced by assistants in interactive development workflows, where context, iteration, tool use, and human oversight differ, remains unresolved.

\subsection{Interpreting the SPDF Rate}
\label{subsec:interpreting_spdf}

The aggregate SPDF rates should be interpreted in light of substantial dataset heterogeneity. RedCode contributed 315 of the 394 candidate file--CWE pairs
(80\%) and 106 of the 120 CONFIRMED exploitations (88\%), so the
aggregate rate is overwhelmingly a property of a single benchmark. The
significant association between dataset source and confirmed
exploitability ($\chi^2(2)=12.55$, $p=0.0019$) reinforces this: SPDF
behaviour is not uniformly distributed across the corpus. Notably,
SecurityEval, the only manually curated, generation-task benchmark of
the three, produced the lowest confirmed rate (5.4\%), whereas the
adversarial and intentionally vulnerable corpora produced the highest
(33.7\% and 28.6\%). The aggregate inclusive rate of roughly one in seven should
therefore be read as an upper-leaning estimate driven by adversarial and
vulnerability-seeded code, not as a direct prediction of how often
statically clean code from realistic generation tasks will prove
exploitable.

A second consideration concerns the shared model backbone across
pipeline stages. Both the CWE Detection Agent and the Dynamic Exploit
Verification Agent are implemented with \texttt{gpt-5.4-nano}, so the
verifier confirms hypotheses generated by a sibling configuration of the
same model. This risks correlated reasoning errors and a degree of
self-confirmation: weaknesses the detector is predisposed to surface may
also be those the verifier is predisposed to demonstrate. The categories
with perfect confirmation rates, CWE-338 and CWE-330, each confirmed in
every candidate instance, are consistent with this concern, since their
verification often reduces to demonstrating that a generator is
non-cryptographic, a property both agents recognize from the same
surface cues. Majority voting over five runs reduces stochastic noise
but cannot break this correlation, because all runs draw on the same
model; pairing a detector and verifier from different model families, or
auditing a stratified subsample against human analysts, would establish
how much of the signal is independent of the model that proposed it.

Finally, the SPDF rate treats every confirmed exploitation as
equivalent, yet the confirmed categories differ markedly in
consequence. The three most frequently confirmed classes, CWE-400
(Uncontrolled Resource Consumption), CWE-338, and CWE-330, are
dominated by denial-of-service and weak-randomness weaknesses, which are
among the easiest to demonstrate at runtime and, in many deployment
contexts, among the lower in direct impact. By contrast, CWE-94 (Code Injection) produced numerous candidate findings but no confirmed exploitations.
Because exploitability and consequence are conflated within a single
rate, the headline figure may overstate the practical severity of the
residual risk, motivating severity-weighted formulations that do not
silently equate \emph{exploitable} with \emph{consequential}.

\subsection{Implications}

For security evaluation pipelines, the results caution against treating
static analysis as the primary measure of risk in AI-generated code, as
is common practice. Static analysis remains an efficient first-line
filter, it eliminated 701 files here at negligible cost; however, it
identifies vulnerable patterns rather than demonstrating consequences,
so the strongest assessments combine it with execution-based
verification rather than relying on either in isolation.

For vulnerability detection, both exploitability and static coverage are
unevenly distributed across weakness classes. Several frequently
confirmed categories were absent from the static-analysis output
entirely (Table~\ref{tab:missed_cwes}), indicating that SPDF behaviour
is not merely isolated false negatives within covered classes but, in
some cases, whole categories falling outside rule-based coverage. These
tend to involve semantic properties, randomness quality, cryptographic
strength, authorization logic, and resource consumption, that resist
syntactic pattern matching, suggesting that reasoning- and
execution-based analysis can complement rule-based tooling.

For agentic security systems, dynamic verification served as both discovery and validation: it confirmed 120 weaknesses that survived earlier stages, while 208 candidate weaknesses were not triggered under the generated harnesses and available execution conditions. These results demonstrate how dynamic verification can distinguish reasoning-based vulnerability hypotheses from vulnerabilities for which concrete runtime evidence can be obtained, without treating unsuccessful exploitation as evidence that the candidate weakness is necessarily absent. This depth is expensive, roughly 96.3 million tokens and 29 hours, so exploit verification is best deployed selectively within a layered strategy in which static analysis narrows the search space and reasoning prioritizes candidates. The broader implication is that exploitability deserves treatment as a first-class security outcome rather than an implicit byproduct of detection.

\section{Threats to Validity} \label{sec:threats}

\subsection{Construct Validity}

The primary construct, runtime exploitability, is an operationalization
rather than a perfect measure of real-world security impact. Because the
sandbox disabled network access and provided no external services,
privileged contexts, or vulnerability-specific dependencies, and because successful verification depends on the agent generating an effective exploit harness, some \texttt{NOT\_TRIGGERED} verdicts and the 46 \texttt{INCONCLUSIVE}
outcomes may reflect environmental limits rather than the absence of a
vulnerability; the reported SPDF rates should therefore be read as a
lower bound on demonstrated exploitability. Two further limitations
apply: the CWE Detection Agent relies on LLM-based reasoning and may
produce both false positives and false negatives, and all verdicts are
generated automatically rather than validated by human security
analysts, so individual outcomes may diverge from expert judgment.

\subsection{Internal Validity}

Several implementation choices may influence the observed rates. Both
agents use \texttt{gpt-5.4-nano}; a different model could change
detection, recognition, and exploit-generation behaviour, and, because
the same model both proposes and confirms weaknesses, a portion of the
confirmation signal may be self-reinforcing, as discussed in
Section~\ref{subsec:interpreting_spdf}. In addition, the Docker
sandbox's execution and resource limits, while improving safety and
reproducibility, may suppress exploits that require system privileges or
external conditions. Finally, majority voting over five runs reduces but
does not eliminate stochastic variability; alternative run counts or
aggregation strategies may yield slightly different estimates.

\subsection{External Validity}

The study examines only Python code from SecurityEval, RedCode, and
CyberNative. Vulnerability patterns, static-analysis effectiveness, and
exploitability may differ for production systems and for languages such
as C/C++, Java, and Rust, so the reported rates should not be read as
universal static-analysis failure rates. The static-pass set also
depends on Bandit v1.8.4.dev21 and Semgrep v1.155.0 under the
\texttt{auto} ruleset with no additional filtering; alternative
versions, rulesets, or commercial scanners would shift it. Because these
corpora emphasize security-relevant tasks, SPDF prevalence in
general-purpose development may differ.

\section{Conclusion and Future Work}

This work examined whether code producing no findings under a composite Bandit–Semgrep static-analysis gate can nevertheless exhibit dynamically demonstrable vulnerabilities. Through the Static-Pass Dynamic-Fail (SPDF) framework, we found that static-analysis outcomes under this evaluated configuration and demonstrated runtime exploitability capture distinct forms of security evidence. More than one in eight samples producing no findings under the evaluated static-analysis gate were subsequently confirmed exploitable under our conservative dynamic-verification criterion, and several CWE categories absent from the Bandit–Semgrep findings were nonetheless confirmed at runtime. Because dynamic verification was restricted to LLM-identified candidates, these values characterize the yield of the evaluated pipeline rather than the total prevalence of exploitable vulnerabilities among all static-pass samples. These results support treating software security assessment as an accumulation of evidence across complementary forms of analysis rather than relying on any single evaluated technique in isolation. Alongside these findings, we
release an agentic exploitability-evaluation framework, a dataset of
dynamically evaluated vulnerability candidates, and a reproducible
methodology for relating static-analysis outcomes to runtime
exploitability.

Two directions follow most directly. First, pairing a detector and
verifier from different model families would test how much of the
confirmed-exploitability signal is independent of the model that
proposed it. Second, weighting confirmed instances by severity would
distinguish exploitable weaknesses from consequential ones, refining the
SPDF rate as a measure of practical risk.

\section*{Acknowledgements}
The authors acknowledge support from the Faculty of Science Undergraduate Research Opportunities program at Toronto Metropolitian University which made this work possible. OpenAI ChatGPT and Grammarly were used to assist with language editing and manuscript refinement. All technical content, experimental design, analysis, and final manuscript decisions were reviewed and validated by the authors.

\bibliographystyle{IEEEtran}
\bibliography{main}

\end{document}